\documentclass[letterpaper,english,a4paper,twocolumn,secnumarabic, amssymb, amsmath,nofootinbib, tightenlines,nobibnotes, showpacs, aps, prd]{revtex4}
\usepackage[T1]{fontenc}
\usepackage[utf8]{inputenc}
\usepackage{amsmath}
\usepackage{amssymb}
\usepackage{graphicx}
\usepackage{esint}

\makeatletter

\providecommand{\tabularnewline}{\\}

\@ifundefined{textcolor}{}
{%
 \definecolor{BLACK}{gray}{0}
 \definecolor{WHITE}{gray}{1}
 \definecolor{RED}{rgb}{1,0,0}
 \definecolor{GREEN}{rgb}{0,1,0}
 \definecolor{BLUE}{rgb}{0,0,1}
 \definecolor{CYAN}{cmyk}{1,0,0,0}
 \definecolor{MAGENTA}{cmyk}{0,1,0,0}
 \definecolor{YELLOW}{cmyk}{0,0,1,0}
}

\usepackage{psfrag}
\makeatletter

\@addtoreset{equation}{section}

\@addtoreset{figure}{section}

\@addtoreset{table}{section}
\makeatother

\usepackage{ifpdf}\ifpdf\usepackage{epstopdf}\usepackage[pdftex,ps2pdf,dvips,colorlinks,urlcolor=blue,citecolor=blue,linkcolor=blue]{hyperref}\else
\usepackage[hypertex,ps2pdf,dvips,colorlinks,urlcolor=blue,citecolor=blue,linkcolor=blue]{hyperref}\fi\pdfadjustspacing=1

\makeatother

\usepackage{babel}
\begin{document}

\title{Diagonal multi-soliton matrix elements in finite volume}

\author{T. Pálmai$^{1}$ and G. Takács$^{1,2}$\\
~\\
$^{1}$Department of Theoretical Physics, \\
Budapest University of Technology and Economics\\
~\\
$^{2}$MTA-BME \textquotedbl{}Momentum\textquotedbl{} Statistical
Field Theory Research Group}

\date{5th February 2013}

\begin{abstract}
We consider diagonal matrix elements of local operators between multi-soliton
states in finite volume in the sine-Gordon model, and formulate a
conjecture regarding their finite size dependence which is valid up
to corrections exponential in the volume. This conjecture extends
the results of Pozsgay and Takács which were only valid for diagonal
scattering. In order to test the conjecture we implement a numerical
renormalization group improved truncated conformal space approach.
The numerical comparisons confirm the conjecture, which is expected
to be valid for general integrable field theories. The conjectured
formula can be used to evaluate finite temperature one-point and two-point
functions using recently developed methods. 
\end{abstract}
\maketitle

\section{Introduction}

In this paper we continue a program to describe form factors in finite
volume, initiated in \cite{Pozsgay:2007kn,Pozsgay:2007gx} and further
developed and extended in the works \cite{Kormos:2007qx,Pozsgay:2008bf,Feher:2011aa,Takacs:2011nb,Feher:2011fn}.
Previously, finite volume form factors have also been studied in other
approaches \cite{Smirnov:1998kv,korepin-slavnov,Mussardo:2003ji}. 

The present effort at describing finite volume form factors is directed
towards an understanding of finite temperature correlation functions
and other related cases where the spectral expansion of a quantity
is ill-defined due to singularities associated with disconnected contributions
present in operator matrix elements. A different attempt to find a
spectral expansion for finite temperature correlators is Doyon's finite
temperature form factor formalism \cite{Doyon:2006pv}.

We work in two-dimensional integrable quantum field theories, where
the $S$ matrix can be obtained from the bootstrap \cite{zam-zam}.
In such theories, using the scattering amplitudes as input it is possible
to obtain a set of equations satisfied by the form factors \cite{Karowski:1978vz}.
The complete system of form factor equations, which provides the basis
for the form factor bootstrap, was proposed in \cite{Kirillov:1987jp}.
For a detailed and thorough exposition of the subject we refer to
\cite{Smirnov:1992vz}. 

Finite volume form factors can be used as a tool to compute finite
temperature correlation functions \cite{Pozsgay:2007gx,Essler:2009zz,Pozsgay:2009pv,Pozsgay:2010cr},
or matrix elements of local fields in a boundary setting \cite{Kormos:2010ae}.
Another application of the formalism is the extension of form factor
perturbation theory, which was developed in \cite{Delfino:1996xp}
to described non-integrable perturbations of integrable quantum field
theories, to higher orders in \cite{Takacs:2009fu}.

One of the applications of our framework is that it allows for a direct
comparison of solutions of the form factor axioms to field theory
dynamics. In fact, this works the other way as well, since we can
use well-established results about the form factor solutions in a
given theory to test our ideas or conjectures that arise in developing
the finite volume form factor program. There is still place for developments,
as we do not yet have a complete description of finite volume form
factors for the case of non-diagonal scattering. The first steps were
taken in \cite{Feher:2011aa} with a study of sine-Gordon breather
and two-soliton form factors; later it was extended to multi-soliton
states \cite{Feher:2011fn}, using the framework for numerical evaluation
of multi-soliton form factors developed in \cite{Palmai:2011kb}.
Another important direction is to incorporate exponential finite size
corrections: so far this was only done for the so-called $\mu$-terms
\cite{Pozsgay:2008bf,Takacs:2011nb}, and even in that case the method
is not yet entirely systematic.

Theories with non-diagonal scattering, in which the spectrum contains
some nontrivial particle multiplets (typically organized into representations
of some group symmetry), such as sine-Gordon or the $O(3)$ nonlinear
sigma model are very important for condensed matter applications (e.g.
to spin chains or one-dimensional electron systems; for a review see
\cite{Essler:2004ht}). The finite volume description of form factors
can be used to develop a low-temperature and large-distance expansion
for finite-temperature correlation functions, which could in turn
be used to explain experimental data, e.g. from inelastic neutron
scattering \cite{Essler:2007jp,Essler:2009zz}. Another interesting
application is to extend the computation of one-point functions of
bulk operators on a strip to the non-diagonal case. Finite volume
methods are also a promising approach in the study of quantum quenches
in integrable quantum field theories \cite{Kormos:2010ae,Schuricht:2012kr}.

In this paper we treat the sine-Gordon model as an example. It can
be considered as the prototype of a non-diagonal scattering theory,
and it has the advantage that its finite volume spectra and form factors
can be studied very effectively numerically using the truncated conformal
space approach (TCSA), originally developed by Yurov and Zamolodchikov
for the scaling Lee-Yang model \cite{Yurov:1989yu}, but later extended
to the sine-Gordon theory \cite{Feverati:1998va}. Its exact form
factors are also known in full generality \cite{Smirnov:1992vz,Lukyanov:1993pn,Lukyanov:1997bp,Babujian:1998uw,Babujian:2001xn},
and so it is a useful playground to test our theoretical ideas on
finite volume form factors. However, before embarking on the present
program, a technical problem had to be solved. Namely, it was clear
from our earlier studies \cite{Feher:2011aa,Feher:2011fn} that truncated
conformal space did not converge very well for diagonal matrix elements.
In order to solve this problem we implemented the numerical renormalization
group (NRG) improvement introduced by Konik and Adamov \cite{Konik:2007cb}.
As we demonstrate, the resulting NRG-TCSA method proved to be accurate
enough to perform a stringent test of our conjectures.

The paper is organized as follows. After a brief review of the necessary
facts about sine-Gordon model and its finite volume soliton form factors
in Section \ref{sec:Solitons-in-finite-volume}, we formulate our
main conjecture in Section \ref{sec:A-conjecture-for-disconnected-parts}.
In Section \ref{sec:Numerical-methods} we give a brief description
of the numerical methods, and then present the results of our computations
in Section \ref{sec:Results}. Section \ref{sec:Conclusions} is reserved
for the conclusions.

\section{Solitons in finite volume\label{sec:Solitons-in-finite-volume}}

\subsection{Action and $S$ matrix\label{sub:Action-and-}}

Sine-Gordon model is defined by the classical action 
\[
\mathcal{A}=\int d^{2}x\left(\frac{1}{2}\partial_{\mu}\Phi\partial^{\mu}\Phi+\frac{m_{0}^{2}}{\beta^{2}}\cos\beta\Phi\right)
\]
The spectrum of the quantum theory is generated by a doublet of a
soliton and an antisoliton, both of mass $M$. Their exact $S$ matrix
can be written as \cite{zam-zam} 
\begin{align}
\mathcal{S}_{i_{1}i_{2}}^{j_{1}j_{2}}(\theta,\xi) & =S_{i_{1}i_{2}}^{j_{1}j_{2}}(\theta,\xi)S_{0}(\theta,\xi)\label{eq:sg_smatrix}\\
 & \xi=\frac{\beta^{2}}{8\pi-\beta^{2}}\nonumber 
\end{align}
where the non-zero elements are
\begin{eqnarray*}
 &  & S_{++}^{++}(\theta,\xi)=S_{--}^{--}(\theta,\xi)=1\\
 &  & S_{+-}^{+-}(\theta,\xi)=S_{-+}^{-+}(\theta,\xi)=S_{T}(\theta,\xi)\\
 &  & S_{+-}^{-+}(\theta,\xi)=S_{-+}^{+-}(\theta,\xi)=S_{R}(\theta,\xi)
\end{eqnarray*}
and

\begin{align*}
S_{T}(\theta,\xi) & =\frac{\sinh\left(\frac{\theta}{\xi}\right)}{\sinh\left(\frac{i\pi-\theta}{\xi}\right)},\quad S_{R}(\theta,\xi)=\frac{i\sin\left(\frac{\pi}{\xi}\right)}{\sinh\left(\frac{i\pi-\theta}{\xi}\right)}\\
S_{0}(\theta,\xi) & =-\exp\left\{ -i\int_{0}^{\infty}\frac{dt}{t}\frac{\sinh\frac{\pi(1-\xi)t}{2}}{\sinh\frac{\pi\xi t}{2}\cosh\frac{\pi t}{2}}\sin\theta t\right\} 
\end{align*}
Besides the solitons, the spectrum of theory contains also breathers
which are bound states of a soliton and antisoliton. 

Sine-Gordon model can also be represented as a free massless boson
conformal field theory (CFT) perturbed by a relevant operator, with
the Hamiltonian
\begin{equation}
H=\int dx\frac{1}{2}:\left(\partial_{t}\Phi\right)^{2}+\left(\partial_{x}\Phi\right)^{2}:+\lambda\int dx:\cos\beta\Phi:\label{eq:pcft_action}
\end{equation}
where the semicolon denotes normal ordering in terms of the modes
of the $\lambda=0$ massless field. Due to anomalous dimension of
the normal ordered cosine operator, the coupling constant $\lambda$
has dimension
\[
\lambda\sim\left[\mbox{mass}\right]^{2-\beta^{2}/4\pi}
\]
so it defines the mass scale of the model and the dimensionless coupling
parameter is $\beta$.

\subsection{Soliton form factors\label{sub:Soliton-form-factors}}

The class of operators we consider consists of exponentials of the
bosonic field $\Phi$. Their vacuum expectation value is known exactly
\cite{Lukyanov:1996jj}:
\begin{eqnarray}
\mathcal{G}_{a}(\beta) & = & \langle\mathrm{e}^{ia\beta\Phi}\rangle=\left[\frac{M\sqrt{\pi}\Gamma\left(\frac{4\pi}{8\pi-\beta^{2}}\right)}{2\Gamma\left(\frac{\beta^{2}/2}{8\pi-\beta^{2}}\right)}\right]^{\frac{a^{2}\beta^{2}}{4\pi}}\label{eq:exactvev}\\
 & \times & \exp\Bigg\{\int_{0}^{\infty}\frac{dt}{t}\Bigg[-\frac{a^{2}\beta^{2}}{4\pi}e^{-2t}\nonumber \\
 &  & +\frac{\sinh^{2}\left(\frac{a}{4\pi}t\right)}{2\sinh\left(\frac{\beta^{2}}{8\pi}t\right)\cosh\left(\left(1-\frac{\beta^{2}}{8\pi}\right)t\right)\sinh t}\Bigg]\Bigg\}\nonumber 
\end{eqnarray}
with $M$ denoting the soliton mass related to the coupling $\lambda$,
defined in (\ref{eq:pcft_action}), via \cite{Zamolodchikov:1995xk}
\begin{equation}
\lambda=\frac{2\Gamma(\Delta)}{\pi\Gamma(1-\Delta)}\left(\frac{\sqrt{\pi}\Gamma\left(\frac{1}{2-2\Delta}\right)M}{2\Gamma\left(\frac{\Delta}{2-2\Delta}\right)}\right)^{2-2\Delta},\quad\Delta=\frac{\beta^{2}}{8\pi}\label{eq:mass_scale}
\end{equation}
Multi-soliton form factors, i.e.
\begin{equation}
F_{i_{1}\dots i_{N}}^{\mathcal{O}}(\theta_{1},\dots,\theta_{N})=\langle0|\mathcal{O}|A_{i_{N}}\left(\theta_{N}\right)\ldots A_{i_{1}}\left(\theta_{1}\right)\rangle,
\end{equation}
($A_{\pm}$ denoting a soliton/antisoliton and the ordering of rapidities
$\theta_{1}>\ldots>\theta_{N}$) have been constructed using several
different approaches: the earliest construction is by Smirnov (reviewed
in \cite{Smirnov:1992vz}), then a free field representation by Lukyanov
\cite{Lukyanov:1993pn,Lukyanov:1997bp}, and later in the work by
Babujian et al. \cite{Babujian:1998uw,Babujian:2001xn}. Here we use
formulae from Lukyanov's work \cite{Lukyanov:1997bp} (in conjunction
with its numerical evaluation method given in \cite{Palmai:2011kb});
however, certain of his conventions are different and therefore we
change the labeling of the form factors accordingly (see eqn. (\ref{eq:fromlukyanovstoours})
below). The reason is that the form factors we use satisfy form factor
bootstrap relations which are slightly different from Lukyanov's conventions;
in this we conform to the conventions of the papers \cite{Pozsgay:2007kn,Pozsgay:2007gx}.
In our notations, the form factor equations are:

I. Lorentz-invariance
\begin{equation}
F_{i_{1}\dots i_{N}}^{\mathcal{O}}(\theta_{1}+\Lambda,\dots,\theta_{N}+\Lambda)=\mathrm{e}^{s(\mathcal{O})\Lambda}F_{i_{1}\dots i_{N}}^{\mathcal{O}}(\theta_{1},\dots,\theta_{N})\label{eq:Lorentzaxiom}
\end{equation}
where $s(\mathcal{O})$ is the Lorentz spin of the operator $\mathcal{O}$.

II. Exchange:
\begin{eqnarray}
 &  & F_{i_{1}\dots i_{k}i_{k+1}\dots i_{N}}^{\mathcal{O}}(\theta_{1},\dots,\theta_{k},\theta_{k+1},\dots,\theta_{N})=\nonumber \\
 &  & \quad S_{i_{k}i_{k+1}}^{j_{k}j_{k+1}}(\theta_{k}-\theta_{k+1})\nonumber \\
 &  & \quad\times F_{i_{1}\dots j_{k+1}j_{k}\dots i_{N}}^{\mathcal{O}}(\theta_{1},\dots,\theta_{k+1},\theta_{k},\dots,\theta_{N})\label{eq:exchangeaxiom}
\end{eqnarray}

III. Cyclic permutation: 
\begin{multline}
F_{i_{1}i_{2}\dots i_{N}}^{\mathcal{O}}(\theta_{1}+2i\pi,\theta_{2},\dots,\theta_{N})=\\
\mathrm{e}^{2\pi i\omega(\mathcal{O})}F_{i_{2}\dots i_{N}i_{1}}^{\mathcal{O}}(\theta_{2},\dots,\theta_{N},\theta_{1})\label{eq:cyclicaxiom}
\end{multline}
where $\omega(\mathcal{O})$ is the mutual locality index between
the operator $\mathcal{O}$ and the asymptotic field that creates
the solitons.

IV. Kinematical singularity
\begin{multline}
-i\mathop{\textrm{Res}}_{\theta=\theta^{'}}F_{i\, k\, i_{1}\dots i_{N}}^{\mathcal{O}}(\theta+i\pi,\theta^{'},\theta_{1},\dots,\theta_{n})=\\
C_{ik'}\left(\delta_{k}^{k'}-\mathrm{e}^{2\pi i\omega(\mathcal{O})}S_{ki_{1}}^{k_{1}j_{1}}(\theta'-\theta_{1})S_{k_{1}i_{2}}^{k_{2}j_{2}}(\theta'-\theta_{2})\right.\\
\left.\dots S_{k_{n-1}i_{n}}^{k'j_{n}}(\theta'-\theta_{N})\right)F_{j_{1}\dots j_{N}}^{\mathcal{O}}(\theta_{1},\dots,\theta_{N})\label{eq:kinematicalaxiom}
\end{multline}
where $C$ is the charge conjugation matrix.

There is a further equation that relates form factors containing breathers
to those containing only solitons, but it is not needed for multi-soliton
states. These equations are supplemented by the assumption of maximum
analyticity (i.e. that the form factors are meromorphic functions
which only have the singularities prescribed by the axioms) and possible
further conditions expressing properties of the particular operator
whose form factors are sought.

The form factors of the operator 
\[
\mathcal{O}_{a}=\mathrm{e}^{ia\beta\Phi}
\]
which satisfy equations (\ref{eq:Lorentzaxiom}-\ref{eq:kinematicalaxiom})
with the locality index
\[
\omega(\mathcal{O}_{a})=a\bmod1
\]
can be obtained from 
\begin{align}
F_{\sigma_{1}\dots\sigma_{2n}}^{a}(\theta_{1},\dots,\theta_{2n}) & =(-1)^{n}\mathcal{F}_{-\sigma_{2N}\dots-\sigma_{1}}^{(a)}(\theta_{2n},\dots,\theta_{1})\nonumber \\
 & =(-1)^{n}\mathcal{F}_{\sigma_{2n}\dots\sigma_{1}}^{(-a)}(\theta_{2n},\dots,\theta_{1})\label{eq:fromlukyanovstoours}
\end{align}
where the functions $\mathcal{F}$ (originally derived by Lukyanov
in \cite{Lukyanov:1997bp}) are specified in appendix A of \cite{Feher:2011fn}.

\subsection{Soliton form factors in finite volume \label{sec:Soliton-form-factors-in-finite-volume}}

The formulae for finite volume form factors, derived in \cite{Pozsgay:2007kn,Pozsgay:2007gx},
were generalized for the case of non-diagonal theories in \cite{Feher:2011aa}
and further investigated in \cite{Feher:2011fn}. Here we only recall
the necessary facts; for more details the reader is referred to the
original papers. 

In finite volume $L$, the space of multi-soliton states can be labeled
by momentum quantum numbers $I_{1},\dots,I_{N}$. We introduce the
following notation for them: 
\begin{equation}
|\{I_{1},I_{2},\dots,I_{N}\}\rangle_{L}^{(r)}\label{eq:finvolstate}
\end{equation}
where the index $r$ enumerates the eigenvectors of the $n$-soliton
transfer matrix, which can be written as 
\begin{multline*}
\mathcal{\mathcal{T}}\left(\vartheta|\left\{ \theta_{1},\dots,\theta_{N}\right\} \right)_{i_{1}\dots i_{N}}^{j_{1}\dots j_{N}}=\\
\mathcal{S}_{ai_{1}}^{c_{1}j_{1}}(\vartheta-\theta_{1})\mathcal{S}_{c_{1}i_{2}}^{c_{2}j_{2}}(\vartheta-\theta_{2})\dots\mathcal{S}_{c_{N-1}i_{N}}^{aj_{N}}(\vartheta-\theta_{N})
\end{multline*}
where $\theta_{1},\dots,\theta_{N}$ are particle rapidities. The
transfer matrix can be diagonalized simultaneously for all values
of $\vartheta$:
\begin{multline*}
\mathcal{\mathcal{T}}\left(\vartheta|\left\{ \theta_{1},\dots,\theta_{N}\right\} \right)_{i_{1}\dots i_{N}}^{j_{1}\dots j_{N}}\Psi_{j_{1}\dots j_{n}}^{(r)}\left(\left\{ \theta_{k}\right\} \right)=\\
t^{(r)}\left(\vartheta,\left\{ \theta_{k}\right\} \right)\Psi_{i_{1}\dots i_{n}}^{(r)}\left(\left\{ \theta_{k}\right\} \right)
\end{multline*}
We can assume that the wave function amplitudes $\Psi^{(r)}$ are
normalized and form a complete basis:
\begin{align}
\sum_{i_{1}\dots i_{N}}\Psi_{i_{1}\dots i_{N}}^{(r)}\left(\left\{ \theta_{k}\right\} \right)\Psi_{i_{1}\dots i_{N}}^{(s)}\left(\left\{ \theta_{k}\right\} \right)^{*} & =\delta_{rs}\label{eq:polarization_normalization}\\
\sum_{r}\Psi_{i_{1}\dots i_{N}}^{(r)}\left(\left\{ \theta_{k}\right\} \right)\Psi_{j_{1}\dots j_{N}}^{(r)}\left(\left\{ \theta_{k}\right\} \right)^{*} & =\delta_{i_{1}j_{1}}\dots\delta_{i_{N}j_{N}}\nonumber 
\end{align}
these eigenfunctions describe the possible polarizations of the $N$
particle state with rapidities $\theta_{1},\dots,\theta_{N}$ inside
the $2^{N}$ dimensional internal space indexed by $i_{1}\dots i_{N}$.
The transfer matrix can be diagonalized using the algebraic Bethe
Ansatz (cf. Appendix A of \cite{Feher:2011aa}), which enables one
to compute the exact form of eigenvalues $t^{(r)}$ and eigenvectors
$\Psi^{(r)}$.

The rapidities of the particles in the state (\ref{eq:finvolstate})
can be determined by solving the quantization conditions
\begin{multline}
Q_{j}^{(r)}(\theta_{1},\dots,\theta_{n})=\\
ML\sinh\theta_{j}+\delta_{j}^{(r)}(\theta_{1},\dots,\theta_{N})=2\pi I_{j},\quad j=1,\dots,N\\
\delta_{j}^{(r)}(\theta_{1},\dots,\theta_{N})=-i\log t^{(r)}\left(\theta_{j},\left\{ \theta_{k}\right\} \right)\label{eq:betheyang_general}
\end{multline}
 When considering rapidities which solve these equations with given
quantum numbers $I_{1},\dots I_{N}$ and a specific polarization state
$r$, they will be written with a tilde as $\tilde{\theta}_{1},\dots,\tilde{\theta}_{N}$.

Using the above ingredients, the finite volume matrix elements can
then be written as \cite{Feher:2011aa,Feher:2011fn}
\begin{multline}
\left|\,^{(s)}\langle\{I_{1}',\dots,I_{M}'\}\vert\mathcal{O}(0,0)\vert\{I_{1},\dots,I_{N}\}\rangle_{L}^{(r)}\right|=\\
\left|\frac{{\displaystyle F^{\mathcal{O}(s)}(\tilde{\theta}_{M}',\dots,\tilde{\theta}_{1}'|\tilde{\theta}_{1},\dots,\tilde{\theta}_{N})^{(r)}}}{\sqrt{\rho^{(r)}(\tilde{\theta}_{1},\dots,\tilde{\theta}_{N})\rho^{(s)}(\tilde{\theta}_{1}',\dots,\tilde{\theta}_{M}')}}\right|+O(\mathrm{e}^{-\mu L})\label{eq:nondiag_genffrelation}
\end{multline}
 where $\rho^{(r)}$ and $\rho^{(s)}$ denote the density of states
of types $r$ and $s$, which can be calculated as the Jacobi determinant
of the Bethe-Yang equations (\ref{eq:betheyang_general}), considering
them as a mapping from the rapidity to quantum number space:
\begin{equation}
\rho^{(r)}(\theta_{1},\dots,\theta_{N})=\det\left\{ \frac{\partial Q_{j}^{(r)}}{\partial\theta_{k}}\right\} _{j,k=1,\dots,N}\label{eq:BY_Jacobian}
\end{equation}
Furthermore,
\begin{eqnarray}
 &  & F^{\mathcal{O}(s)}(\theta_{M}',\dots,\theta_{1}'|\theta_{1},\dots,\theta_{N})^{(r)}=\nonumber \\
 &  & \,\sum_{j_{1}\dots j_{M}}\sum_{i_{1}\dots i_{N}}\Psi_{j_{1}\dots j_{M}}^{(s)}\left(\left\{ \theta_{k}'\right\} \right)^{*}\nonumber \\
 &  & \quad\times F_{\bar{j}_{M}\dots\bar{j}_{1}i_{1}\dots i_{N}}^{\mathcal{O}}(\theta_{M}'+i\pi,\dots,\theta_{1}'+i\pi,\theta_{1},\dots,\theta_{N})\nonumber \\
 &  & \qquad\qquad\qquad\times\Psi_{i_{1}\dots i_{N}}^{(r)}\left(\left\{ \theta_{k}\right\} \right)\label{eq:polarizedff}
\end{eqnarray}
is the $(s,r)$-polarized form factor (the bar denotes the antiparticle).
The absolute value in (\ref{eq:nondiag_genffrelation}) is necessary
to account for the different phase conventions of the multi-particle
states used in the form factor bootstrap and in the finite volume
calculations.

\section{A conjecture for diagonal matrix elements\label{sec:A-conjecture-for-disconnected-parts}}

Relation (\ref{eq:nondiag_genffrelation}) is only valid for matrix
elements with no disconnected pieces, i.e. when the rapidities in
the two finite volume states are all different from each other. If
there are particles with exactly coinciding rapidities in the two
states, i.e. $\tilde{\theta}_{k}'=\tilde{\theta}_{l}$ for some $k$
and $l$, then there are further contributions. Note that equality
of two quantum numbers such as $I_{k}'=I_{l}$ is not sufficient for
the presence of a disconnected contribution, as the corresponding
rapidities will in general be different due to the terms involving
the phase shifts $\delta_{j}^{(r)}$. Therefore such terms are only
present for the case when the two sets of quantum numbers are exactly
identical, and also in the special case when the two states each contain
a particle with exactly zero rapidity. At present, the disconnected
terms are only known for states with diagonal scattering; the form
of these contributions was obtained in \cite{Pozsgay:2007gx}.

Here we present a conjecture for diagonal matrix elements with non-diagonal
scattering, which is an educated guess based on the results valid
for diagonal scattering and also on some lessons learned from nested
Bethe Ansatz systems \cite{Pozsgay:2012kk,Pozsgay:2012kl}. For diagonal
scattering, the formula introduced in \cite{Pozsgay:2007gx} states
that the diagonal finite volume matrix element can be computed as
\begin{multline}
\langle\{I_{1},\dots,I_{N}\}\vert\mathcal{O}(0,0)\vert\{I_{1},\dots,I_{N}\}\rangle_{L}=\\
\frac{1}{\rho(\{1,\dots,N\})_{L}}\sum_{A\subset\{1,2,\dots N\}}\mathcal{F}(A)_{L}\rho(\bar{A})_{L}+O(\mathrm{e}^{-\mu L})\label{eq:diaggenrule}
\end{multline}
where $\bar{A}=\{1,2,\dots N\}\setminus A$. Let us denote the elements
of the sets $A$ and $\bar{A}$ as
\begin{align*}
A & =\{A_{1},\dots,A_{l}\}\\
\bar{A} & =\{\bar{A}_{1},\dots,\bar{A}_{N-l}\}
\end{align*}
with $l=|A|$ the cardinal number (number of elements) of the set
$A$. Then 
\[
\rho(\bar{A})_{L}=\rho(\tilde{\theta}_{\bar{A}_{1}},\dots,\tilde{\theta}_{\bar{A}_{N-l}})_{L}
\]
is the $l$-particle Bethe-Yang Jacobi determinant involving only
the subset $\bar{A}$ of the $N$ particles, and
\[
\mathcal{F}(A)_{L}=F_{l}^{s}(\tilde{\theta}_{A_{1}},\dots,\tilde{\theta}_{A_{l}})
\]
where
\begin{multline*}
F_{l}^{s}(\theta_{1},\dots,\theta_{l})_{i_{1}\dots i_{l}}=\\
\lim_{\epsilon\rightarrow0}F^{\mathcal{O}}(\theta_{l}+i\pi+\epsilon,\dots,\theta_{1}+i\pi+\epsilon,\theta_{1},\dots,\theta_{l})_{\bar{i_{l}}\dots\bar{i_{1}}i_{1}\dots i_{l}}
\end{multline*}
is the so-called symmetric evaluation of the diagonal form factor
involving the particles in set $A$ (the bar denotes the antiparticle).

The main observation is that the density $\rho$ is only well-defined
for states that have an internal state which is an eigenvector of
the multi-soliton transfer matrix. Therefore, in order to define the
disconnected term corresponding to a given subset $A$, the $N$-soliton
wave function amplitude must be decomposed accordingly. Let us suppose
that 
\[
|\{I_{1},I_{2},\dots,I_{N}\}\rangle_{L}^{(r)}
\]
is a finite volume state which corresponds to a wave function eigenvector
\[
\Psi_{i_{1}\dots i_{N}}^{(r)}\left(\left\{ \theta_{k}\right\} \right)
\]
Using the orthogonality and completeness of the $\Psi$ amplitudes
(eqn. (\ref{eq:polarization_normalization})), for any subset $A\subset\{1,2,\dots N\}$
we can define appropriate branching coefficients $\mathcal{C}$ to
split the wave function into pieces given by tensor products of transfer
matrix eigenvectors for the sets $A$ and $\bar{A}$ as follows:
\begin{multline*}
\Psi_{i_{1}\dots i_{N}}^{(r)}\left(\left\{ \theta_{k}\right\} \right)=\sum_{s,t}\mathcal{C}_{st}^{(r)}\left(\left\{ \theta_{k}\right\} |A\right)\\
\times\Psi_{\{i_{k}\}_{k\in A}}^{(s)}\left(\theta_{A_{1}},\dots,\theta_{A_{l}}\right)\Psi_{\{i_{k}\}_{k\in\bar{A}}}^{(t)}\left(\left\{ \theta_{k}\right\} _{k\in\bar{A}}\right)
\end{multline*}
where the sum over $s$ runs over all possible polarization states
(transfer matrix eigenvectors) for $l$ particles, and the sum over
$t$ runs similarly for polarization states of $N-l$ particles. From
(\ref{eq:polarization_normalization}) it follows that the branching
coefficients are normalized as
\[
\sum_{s,t}\left|\mathcal{C}_{st}^{(r)}\left(\left\{ \theta_{k}\right\} |A\right)\right|^{2}=1
\]
The conjectured generalization of (\ref{eq:diaggenrule}) is then
\begin{eqnarray}
 &  & \,^{(r)}\langle\{I_{1},\dots,I_{N}\}\vert\mathcal{O}(0,0)\vert\{I_{1},\dots,I_{N}\}\rangle_{L}^{(r)}=\nonumber \\
 &  & \quad\frac{1}{\rho^{(r)}(\{1,\dots,N\})_{L}}\sum_{A\subset\{1,2,\dots N\}}\label{eq:nondiag_diaggenrule}\\
 &  & \quad\sum_{s,t}\left|\mathcal{C}_{st}^{(r)}\left(\left\{ \tilde{\theta}_{k}\right\} |A\right)\right|^{2}\mathcal{F}^{(s)}(A)_{L}\rho^{(t)}(\bar{A})_{L}+O(\mathrm{e}^{-\mu L})\nonumber 
\end{eqnarray}
where 
\[
\mathcal{F}^{(s)}(A)_{L}=\lim_{\epsilon\rightarrow0}F^{\mathcal{O}(s)}(\tilde{\theta}_{A_{l}}+\epsilon,\dots,\tilde{\theta}_{A_{1}}+\epsilon|\tilde{\theta}_{A_{1}},\dots,\tilde{\theta}_{A_{l}})^{(s)}
\]
is the symmetric diagonal limit of the $(s,s)$-polarized form factor
from (\ref{eq:polarizedff}) and 
\[
\rho^{(t)}(\bar{A})_{L}=\rho^{(t)}(\tilde{\theta}_{\bar{A}_{1}},\dots,\tilde{\theta}_{\bar{A}_{N-l}})
\]
is the density of states in the $\Psi^{(t)}$ channel. 

It is easy to see that eqn. (\ref{eq:nondiag_diaggenrule}) reproduces
(\ref{eq:diaggenrule}) for diagonal scattering, as in that case there
is only a single polarization state for any given number of particles
and the branching coefficients $\mathcal{C}$ are all equal to $1$.
Indeed the structure of (\ref{eq:nondiag_diaggenrule}) is very obvious,
the only ambiguity is which evaluation of the diagonal form factor
to use; for that we substituted the symmetric one, to keep the correspondence
with (\ref{eq:diaggenrule}). 

The first nontrivial case arises when $N=2$. In this subspace there
are two states with non-diagonal scattering, which contain a soliton
and an antisoliton. The wave function amplitudes of the transfer matrix
eigenstates are
\[
\Psi^{(+)}=\frac{1}{\sqrt{2}}\left(\begin{array}{c}
0\\
1\\
1\\
0
\end{array}\right)\qquad\Psi^{(-)}=\frac{1}{\sqrt{2}}\left(\begin{array}{c}
0\\
1\\
-1\\
0
\end{array}\right)
\]
which are nothing else but the branching coefficients of these states
in a (one-soliton)$\times$(one-soliton) basis; it is a great simplifying
feature that these are rapidity independent. The general formula (\ref{eq:nondiag_diaggenrule})
specified to this case yields
\begin{multline}
\,^{(\pm)}\langle\{I_{1},I_{2}\}\vert\mathcal{O}(0,0)\vert\{I_{1},I_{2}\}\rangle_{L}^{(\pm)}=\\
\frac{1}{\rho^{(\pm)}(\tilde{\theta}_{1},\tilde{\theta}_{2})}\bigg[\lim_{\epsilon\rightarrow0}F_{4}^{\mathcal{O}(\pm)}(\tilde{\theta}_{2}+\epsilon,\tilde{\theta}_{1}+\epsilon|\tilde{\theta}_{1},\tilde{\theta}_{2})^{(\pm)}+\\
F_{2s}^{\mathcal{O}}\left(\rho_{1}(\tilde{\theta}_{1})+\rho_{1}(\tilde{\theta}_{2})\right)+\rho^{(\pm)}(\tilde{\theta}_{1},\tilde{\theta}_{2})\left\langle \mathcal{O}\right\rangle \bigg]\label{eq:ssbar_finvol}
\end{multline}
where 
\[
\rho_{1}(\theta)=ML\cosh\theta
\]
is the one-soliton state density, the diagonal one-soliton form factor
is 
\[
F_{2s}^{\mathcal{O}}=F_{+-}^{\mathcal{O}}(\theta+i\pi,\theta)=F_{-+}^{\mathcal{O}}(\theta+i\pi,\theta)
\]
and it is independent of $\theta$ due to Lorentz invariance. In addition,
the $(\pm,\pm)$-polarized two-soliton--two-soliton form factors are
given by 
\begin{multline*}
F_{4}^{\mathcal{O}(s)}(\theta_{2}',\theta_{1}'|\theta_{1},\theta_{2})^{(r)}=\\
\frac{1}{2}\Bigg[F_{+-+-}^{\mathcal{O}}(\theta_{2}'+i\pi,\theta_{1}'+i\pi,\theta_{1},\theta_{2})\\
+rF_{+--+}^{\mathcal{O}}(\theta_{2}'+i\pi,\theta_{1}'+i\pi,\theta_{1},\theta_{2})\\
+sF_{-++-}^{\mathcal{O}}(\theta_{2}'+i\pi,\theta_{1}'+i\pi,\theta_{1},\theta_{2})\\
+rsF_{-+-+}^{\mathcal{O}}(\theta_{2}'+i\pi,\theta_{1}'+i\pi,\theta_{1},\theta_{2})\Bigg]
\end{multline*}
with $r,s=\pm1$. The rapidities $\tilde{\theta}_{1,2}$ can be obtained
by solving the quantization conditions:
\begin{align}
Q_{1}^{(\pm)}(\theta_{1},\theta_{2}) & =ML\sinh\theta_{1}+\delta_{\pm}(\theta_{1}-\theta_{2})=2\pi I_{1}\nonumber \\
Q_{2}^{(\pm)}(\theta_{1},\theta_{2}) & =ML\sinh\theta_{2}+\mathcal{\delta}_{\pm}(\theta_{2}-\theta_{1})=2\pi I_{2}\label{eq:ssbarby}
\end{align}
where the phase-shifts $\delta_{\pm}$ are defined from the eigenvalues
of the two-soliton $S$-matrix in the neutral subspace by 
\begin{align*}
\mathcal{S}_{+}(\theta) & =\mathcal{S}_{+-}^{+-}(\theta)+\mathcal{S}_{+-}^{-+}(\theta)=-\mathrm{e}^{i\delta_{+}(\theta)}\\
\mathcal{S}_{-}(\theta) & =\mathcal{S}_{+-}^{+-}(\theta)-\mathcal{S}_{+-}^{-+}(\theta)=\mathrm{e}^{i\delta_{-}(\theta)}
\end{align*}
where the minus sign introduced in the first line is a redefinition,
which ensures that the phase-shifts are odd and continuous functions
of the rapidity $\theta$. Due to this convention the even states
are quantized with half-integer, while the odd states are quantized
with integer quantum numbers. The densities $\rho^{(\pm)}$ of neutral
two-soliton states can be written as the Jacobi determinant \cite{Feher:2011aa}
\[
\rho^{(\pm)}(\theta_{1},\theta_{2})=\left|\begin{array}{cc}
\frac{\partial Q_{1}^{(\pm)}}{\partial\theta_{1}} & \frac{\partial Q_{1}^{(\pm)}}{\partial\theta_{2}}\\
\frac{\partial Q_{2}^{(\pm)}}{\partial\theta_{1}} & \frac{\partial Q_{2}^{(\pm)}}{\partial\theta_{2}}
\end{array}\right|
\]

\section{Numerical methods \label{sec:Numerical-methods}}

\subsection{TCSA in a nutshell}

To evaluate the form factors numerically, we use the truncated conformal
space approach (TCSA) pioneered by Yurov and Zamolodchikov \cite{Yurov:1989yu}.
The basic idea of this approach is to consider the field theoretic
Hamiltonian on the space of states of the conformal field theory,
and truncate the basis of this space to finitely many vectors by placing
an appropriate upper energy cutoff. The extension to the sine-Gordon
model was developed in \cite{Feverati:1998va} and has found numerous
applications since then. The Hilbert space can be split by the eigenvalues
of the topological charge $\mathcal{Q}$ (or winding number) and the
spatial momentum $P$, where the eigenvalues of the latter are of
the form
\[
\frac{2\pi s}{L}
\]
$s$ is called the 'conformal spin'. In sectors with vanishing topological
charge, we can reduce the size of the Hilbert space using the symmetry
of the Hamiltonian under conjugation of the solitonic charge: 
\[
\mathcal{C}:\qquad\Phi(x,t)\rightarrow-\Phi(x,t)
\]
To fix our notations, we give here the formal definition for the cut-off
Hilbert space:
\begin{multline*}
\mathcal{H}_{s,q}^{(\pm)}(e_{cut})=\\
\left\{ |\Psi\rangle:\;\mathcal{Q}|\Psi\rangle=q|\Psi\rangle,\;(L_{0}-\bar{L}_{0})|\Psi\rangle=s|\Psi\rangle,\;\mathcal{C}|\Psi\rangle=\pm|\Psi\rangle\right.\\
\left.\mbox{and}\quad(L_{0}+\bar{L}_{0}-1/12)|\Psi\rangle=e|\Psi\rangle\quad\mbox{with}\quad e\leq e_{cut}\right\} 
\end{multline*}
where $|\Psi\rangle$ runs over the Hilbert space of the ultraviolet
massless boson with quasi-periodic boundary conditions
\[
\Phi(x+L,t)=\Phi(x,t)+\frac{2\pi}{\beta}q\qquad q\in\mathbb{Z}
\]
$L_{0}$ and $\bar{L}_{0}$ are the usual Virasoro generators, and
$e_{cut}$ is the (dimensionless) cut-off parameter. The upper index
$\pm$ corresponding to projections under $\mathcal{C}$ is applicable
only in the $\mathcal{Q}=0$ sectors.

For the purpose of form factor calculations, we choose the operator
\[
\mathcal{O}=:\mathrm{e}^{i\beta\Phi}:
\]
where the semicolons denote normal ordering with respect to the $\lambda=0$
free massless boson modes. It has the conformal dimensions
\[
\Delta_{\mathcal{O}}=\bar{\Delta}_{\mathcal{O}}=\frac{\beta^{2}}{8\pi}
\]
Using relation (\ref{eq:mass_scale}) we can express all energy levels
and matrix elements in units of (appropriate powers of) the soliton
mass $M$, and we also introduce the dimensionless volume variable
$l=ML$. The general procedure is the same as in \cite{Pozsgay:2007kn,Pozsgay:2007gx}:
the particle content of energy levels can be identified by matching
the numerical TCSA spectrum against the predictions of the Bethe-Yang
equations (\ref{eq:betheyang_general}). After identification, one
can compare the appropriate matrix elements to the theoretical values
given by (\ref{eq:nondiag_genffrelation}) and (\ref{eq:nondiag_diaggenrule}).
There are some technical issues in the identification of states due
to level crossings, which can also affect numerical accuracy; for
a discussion of these we refer to \cite{Kormos:2007qx,Feher:2011aa}.

\subsection{The numerical renormalization group in TCSA}

As mentioned in the introduction, our earlier studies showed \cite{Feher:2011aa,Feher:2011fn}
that TCSA did not converge very well for diagonal matrix elements.
This means that the numerically evaluated matrix elements showed a
marked dependence on the cutoff $e_{cut}$. It is also known from
earlier experience that TCSA converges faster for smaller values of
$\beta$ (or $\xi$), and this was also clear from previous calculations
\cite{Takacs:2005fx,Takacs:2009fu,Feher:2011aa,Feher:2011fn}: decreasing
$\xi$ the cut-off dependence was reduced and the agreement between
the numerical TCSA results and the predictions of the finite volume
form factor formalism was improved at the same time.

Following the proposal of Konik and Adamov \cite{Konik:2007cb}, we
use a Wilsonian type numerical renormalization group (NRG) to improve
the precision of TCSA. This allows us to use much higher values for
the cutoff than for the usual TCSA. Usually, the attainable cutoff
is in the range $e_{cut}=20\dots26$ (depending on the sector and
the value of $\beta$), which corresponds to an upper limit of around
$20000$ states. E.g. in the $\mathcal{Q}=2$, $s=0$ sector at $\xi=\frac{2}{7}$,
$e_{cut}=26$ translates to $23771$ states. With the NRG improvement,
it is possible to take into account the effect of several hundred
thousand states, e.g. in the previous example we could take into account
$840000$ states corresponding to the cutoff value of $38$ (the number
of states increases exponentially with the cutoff). 

Our implementation of the NRG in TCSA is as follows. Let us take a
given value $e_{cut}$ of the cutoff with $N$ states, which we order
by increasing value of their conformal energy eigenvalue. We start
with the lowest lying $N_{0}$ of them (which is chosen to correspond
to some lower value $e_{cut,0}$ of the cutoff ). We split the remaining
states into $k$ ``shells'', each containing $N_{1}$ states and
a ``remainder shell'' of $N_{r}$ states such that 
\[
N=N_{0}+kN_{1}+N_{r}
\]
The idea is that first we diagonalize the lowest $N_{0}\times N_{0}$
block of the Hamiltonian and retain the first $n$ lowest lying eigenvalues
with their corresponding eigenvectors. Then we add the first shell
to the basis and recompute the lowest $n$ eigenvalues and corresponding
eigenvectors, which are now $N_{0}+N_{1}$ dimensional. In the next
step, we add to these modified eigenvectors the next shell, and recompute
the eigenvalues and eigenvectors, which in the original conformal
basis will now have $N_{0}+2N_{1}$ dimensions. Repeating the procedure
$k$ times and applying it to the remainder shell, we arrive at results
for the first $n$ eigenvectors in the $N$ dimensional Hilbert space.

It is obvious that this procedure introduces a new element of approximation,
namely we partially neglect part the matrix elements mixing the higher
shells (part of it is retained as the starting $N_{0}$ basis vectors
at step $l$ will contain components from the previous $l-1$ shells).
In terms of the parameters there are two sources of error: setting
i) $k>1$ and ii) $n\neq N_{0}$. E.g. for $n=N_{0}$ the procedure
would be exact when $k=1$, i.e. if the completion is done in one
step. The error resulting from these sources is difficult to control.
Nevertheless, our experience shows that by choosing the starting dimension
$N_{0}$ and the number of steps $k$ well, the results are very much
improved. For low values of the total dimension $N$, a direct comparison
to the exact spectrum in the truncated space is possible and this
was carried out as a preliminary study to assess the reliability of
the method (see later).

\begin{figure}
\begin{centering}
\includegraphics[scale=0.55]{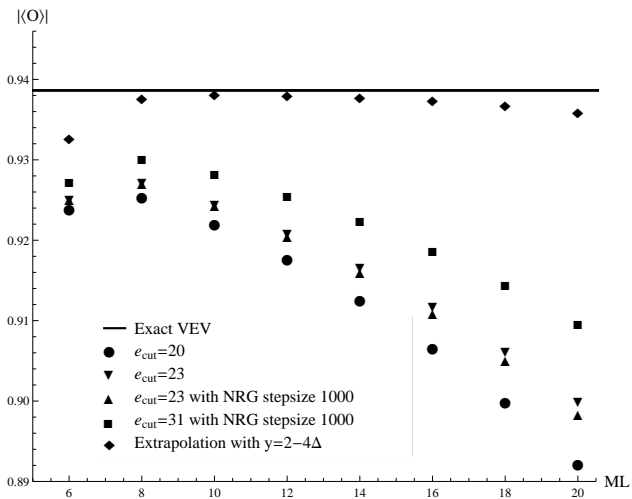}
\par\end{centering}

\caption{\label{fig:vev}TCSA evaluation of the vacuum expectation value at
$\xi=2/7$, with and without NRG, compared to the exact result (\ref{eq:exactvev}).
An extrapolation using (\ref{eq:firstorder_rg}) is also shown. The
horizontal axis is the dimensionless volume $ML$, while the vertical
one is the absolute value of the dimensionless vacuum expectation
value in units of the soliton mass (the true value of both the predicted
and the measured value is negative).}
\end{figure}

Once the cutoff is high enough one can apply a perturbative renormalization
group to extrapolate to infinite cutoff \cite{Feverati:2006ni,Konik:2007cb,Giokas:2011ix,Watts:2011cr}.
To leading order, for every quantity $\mathcal{M}$ there is a renormalization
group exponent $y$ such that the leading cutoff dependence is
\begin{equation}
\mathcal{M}(e_{cut})=\mathcal{M}_{\infty}+A(e_{cut})^{-y}+\dots\label{eq:firstorder_rg}
\end{equation}
where $y$ can be computed using ultraviolet perturbation theory \cite{Konik:2007cb}
which predicts that in our case %
\footnote{G. Watts, private communication. See also the result for the magnetization
operator in the Ising model perturbed by external magnetic field in
\cite{Konik:2007cb}, which is directly analogous to the situation
we have (i.e. the measured operator is the same as the perturbing
one).%
}
\[
y=2-4\Delta
\]
where
\[
\Delta=\frac{\xi}{\xi+1}
\]
is the conformal dimension of both the measured and the perturbing
operator (in our case they are equal). The validity of the extrapolation
with the above exponent can be demonstrated for the the vacuum expectation
value as shown in figure \ref{fig:vev}, and the results in Section
\ref{sec:Results} show that it also works for the diagonal matrix
elements.

In Fig. \ref{fig:vev} numerical evaluations of the vacuum expectation
value at $\xi=\frac{2}{7}$ are displayed. The NRG flow of $|\langle\mathcal{O}\rangle|$
was initiated at $e_{cut,0}=20$ ($5614$ states). For $e_{cut}=23$
($16032$ states) TCSA calculations with and without NRG were also
carried out. The difference between the two sets of data points is
small. Note that using a step size of $100$ states produced data
differing from the displayed NRG-TCSA points only in the $6$th digits.
Results for $e_{cut}=31$ ($181046$ states) and the extrapolation
(\ref{eq:firstorder_rg}) using data up to $e_{cut}=31$ are also
shown. 

In conclusion, utilizing the NRG (and RG) improvement it is possible
to obtain a satisfactory agreement between the theoretical and TCSA
vacuum expectation values. Based on this preliminary study, it is
also reasonable to believe that the NRG error remains small for other
diagonal matrix elements, thus the NRG improved TCSA should be reliable.

\section{Results \label{sec:Results}}

\subsection{Diagonal scattering}

Before presenting the more interesting results corroborating our conjecture
(\ref{eq:nondiag_diaggenrule}) we show some data confirming (\ref{eq:diaggenrule}),
which also serve to test the accuracy of the NRG-TCSA method.

\begin{figure}
\begin{centering}
\begin{tabular}{c}
\includegraphics[scale=0.57]{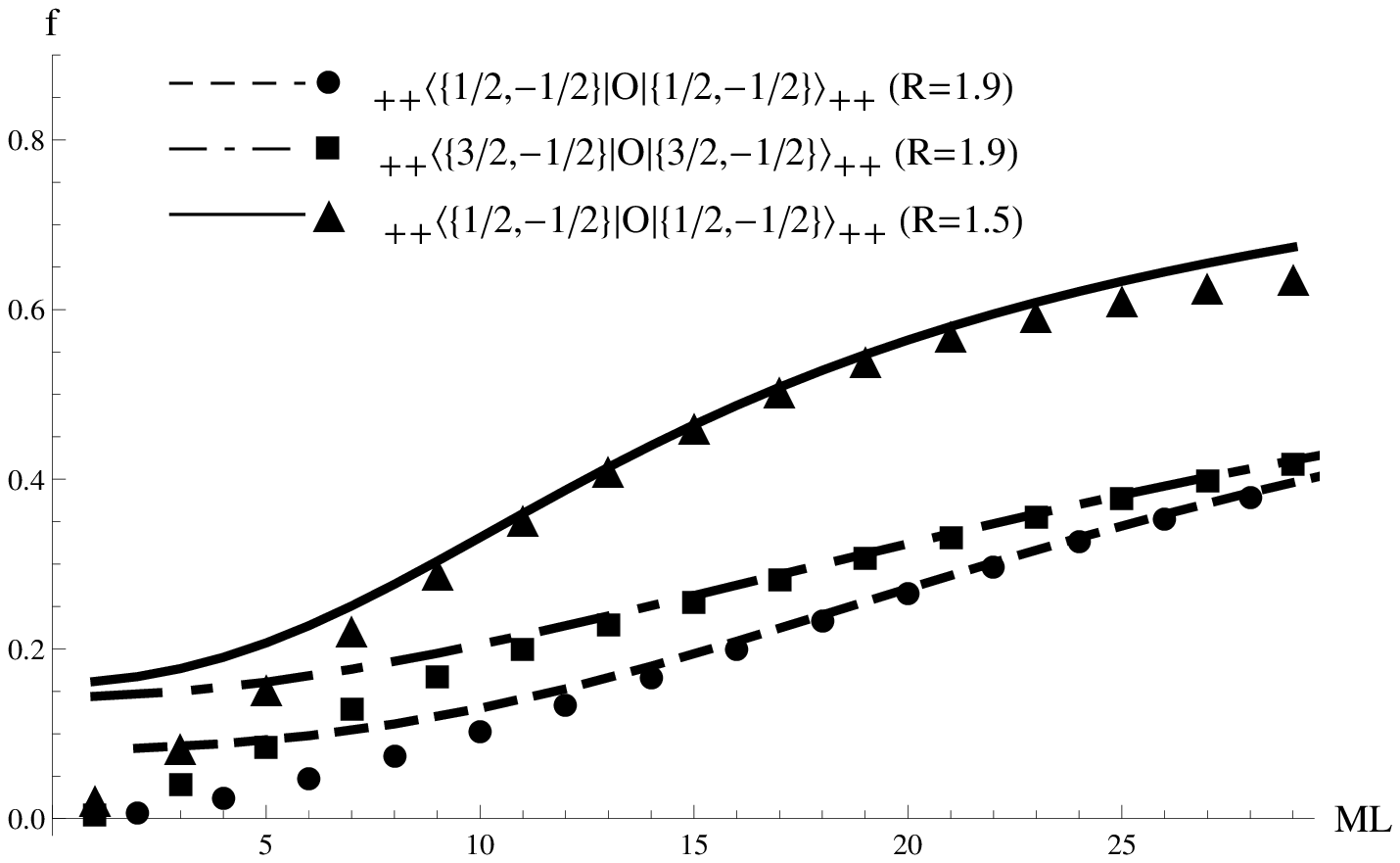}\tabularnewline
(a)\tabularnewline
\includegraphics[scale=0.57]{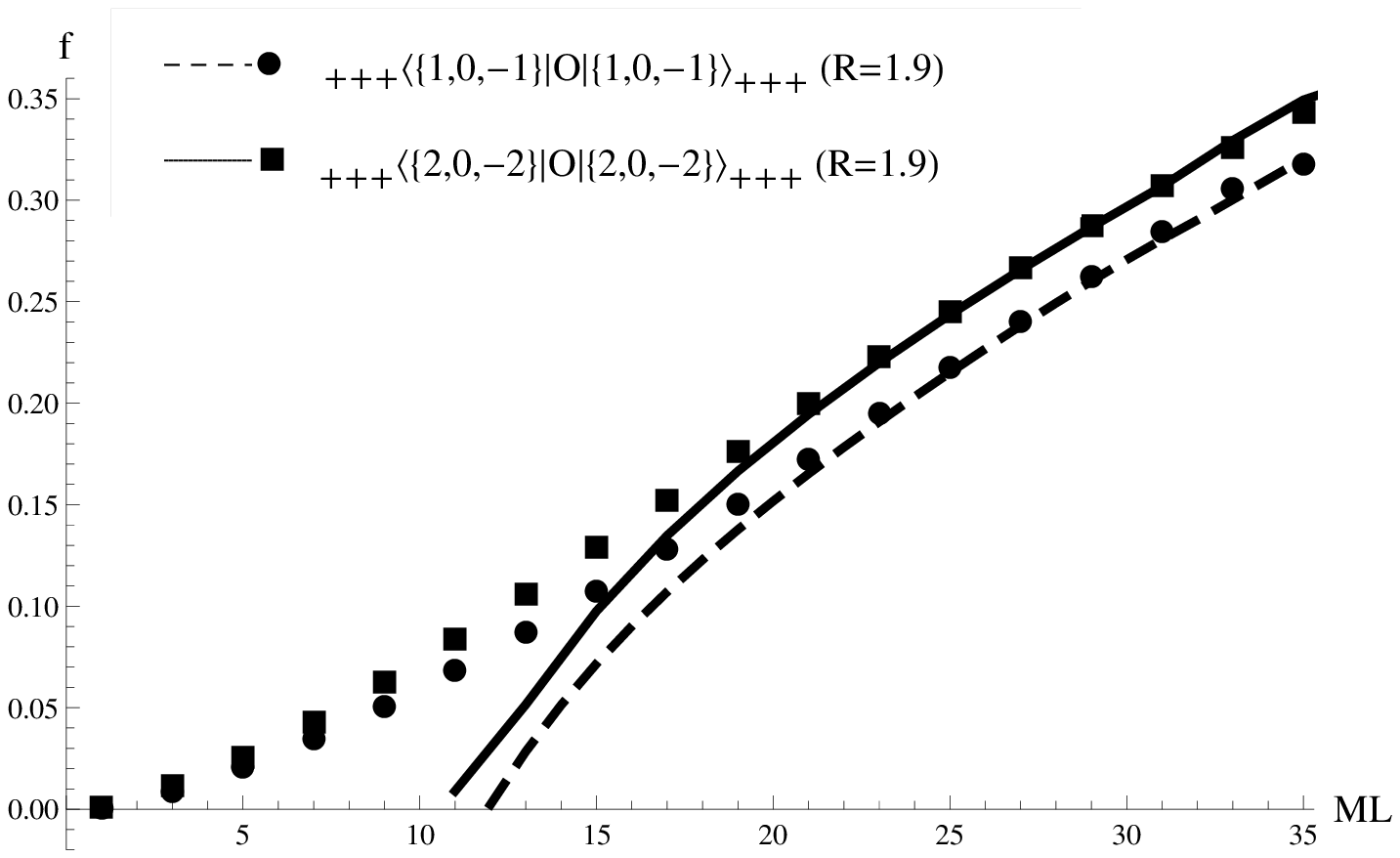}\tabularnewline
(b)\tabularnewline
\end{tabular}
\par\end{centering}

\caption{\label{fig:diag}Diagonal matrix elements with (a) $2$ solitons and
(b) $3$ solitons, all of them of the same ($+$) charge. (Here and
in the subsequent figures $f$ denotes the absolute value of the finite
volume matrix elements in units of $M$.)}
\end{figure}

This case was already investigated in \cite{Feher:2011fn}. However,
in that work we found that the agreement between theory and TCSA results
was rather unsatisfactory, especially for $\xi=2/7$. In contrast,
the NRG greatly improves the situation, as shown in Fig. \ref{fig:diag}
where two and three-soliton diagonal matrix elements corresponding
to the lowest lying states in the appropriate sectors are plotted.
The data are parametrized by the compactification radius of the bosonic
field defined by 
\[
R=\frac{\sqrt{4\pi}}{\beta}
\]
and related to $\xi$ as 
\[
\xi=\frac{1}{2R^{2}-1}
\]
$\xi=2/7$ and $\xi=50/311$ correspond to $R=1.5$ and $1.9$, respectively.
For the two-soliton data we employed the extrapolation (\ref{eq:firstorder_rg})
using NRG-TCSA data up to $e_{cut}=34$ (corresponding to approx.
277 thousand states), while for the three-soliton data the extrapolation
was applied to NRG-TCSA data up to $e_{cut}=37$ (corresponding to
approx. 140 thousand states). The agreement is very good even in the
less trivial three-soliton case, where six-soliton form factors are
encountered on the bootstrap side.

\subsection{Non-diagonal scattering}

\begin{figure*}

\begin{center}
\begin{tabular}{ccc}
\includegraphics[scale=0.55]{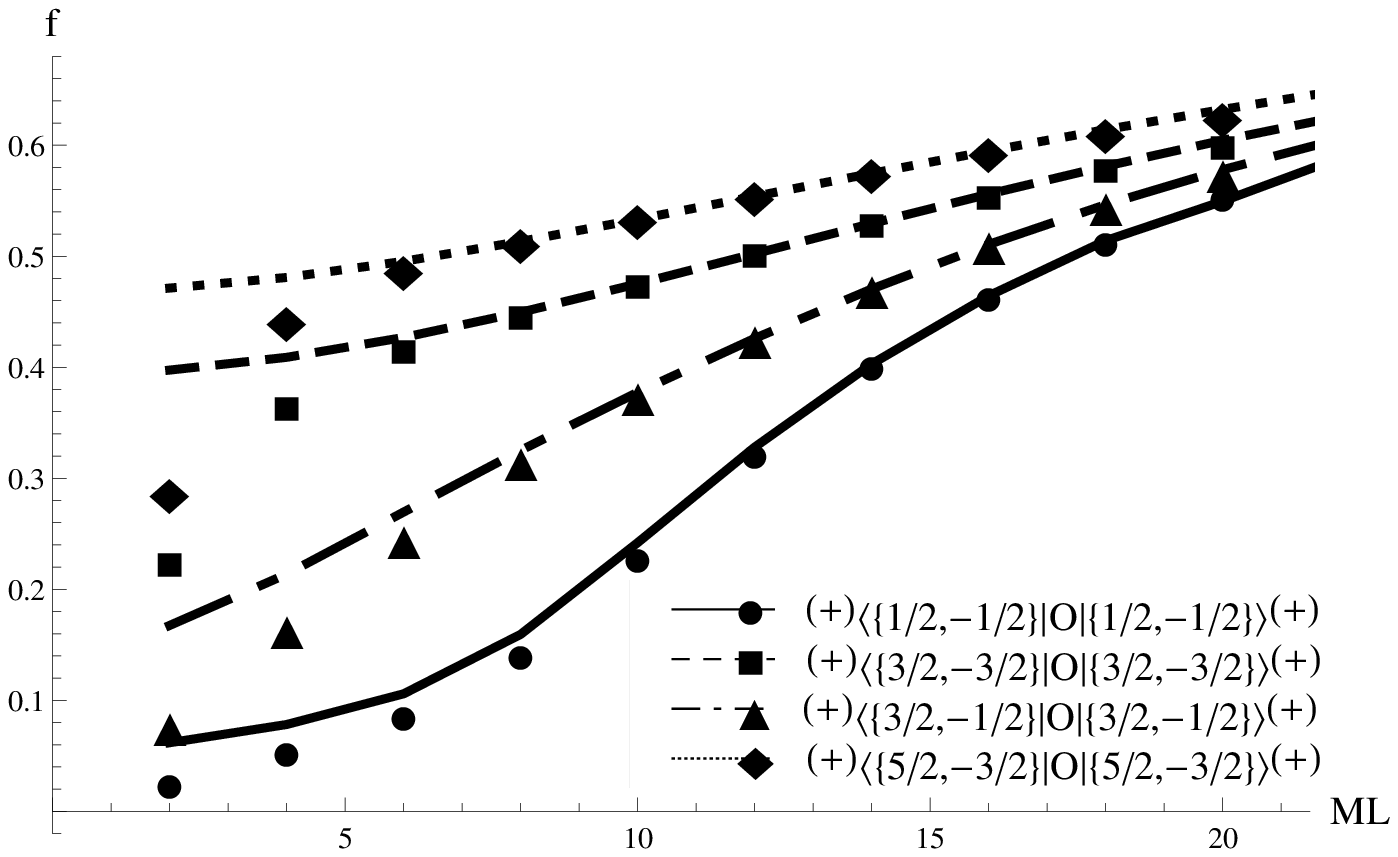} & $\quad$ & \includegraphics[scale=0.55]{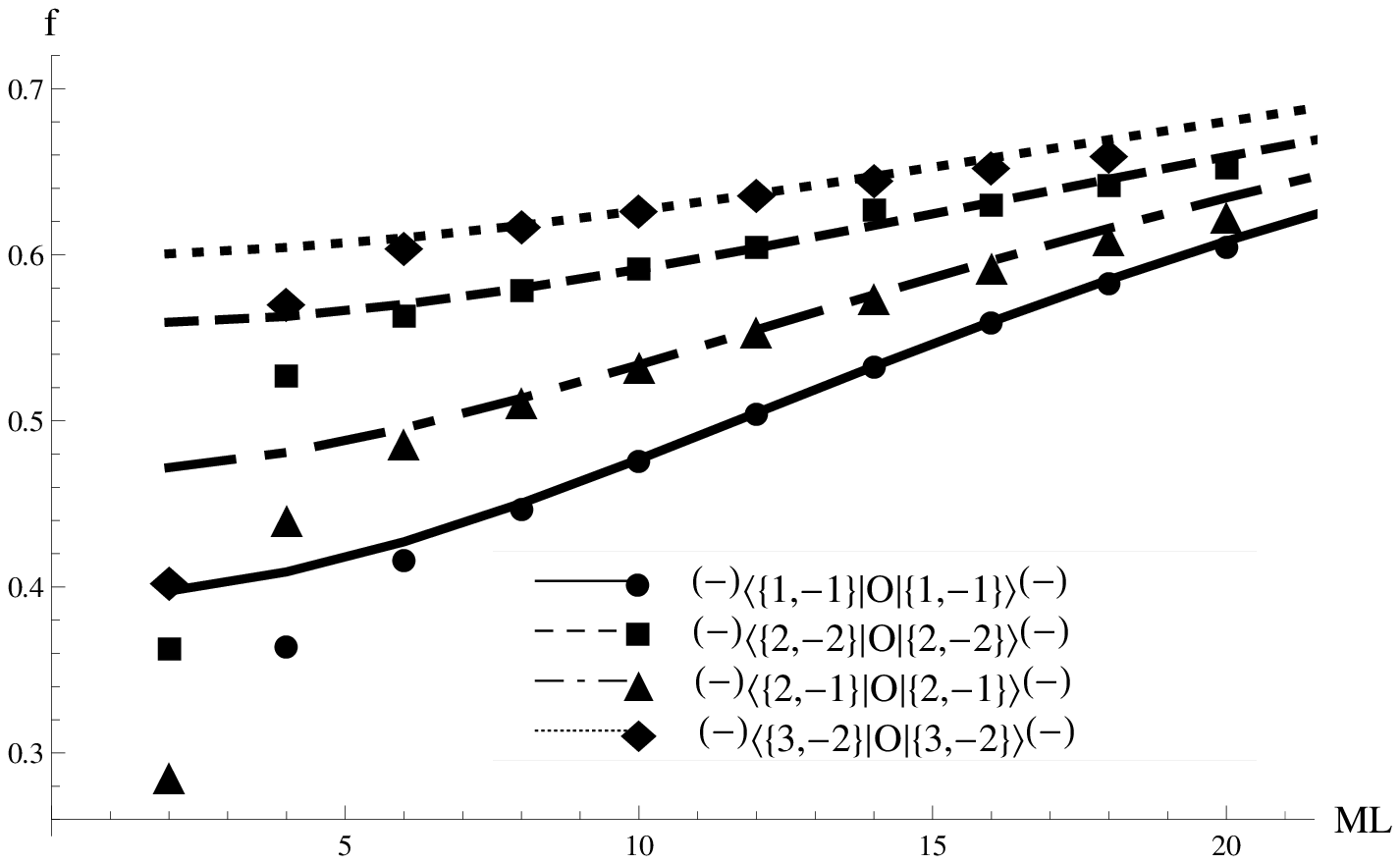}\tabularnewline
$R=1.5$, even states &  & $R=1.5$, odd states\tabularnewline
\end{tabular}
\par\end{center}

\begin{center}
\begin{tabular}{ccc}
\includegraphics[scale=0.55]{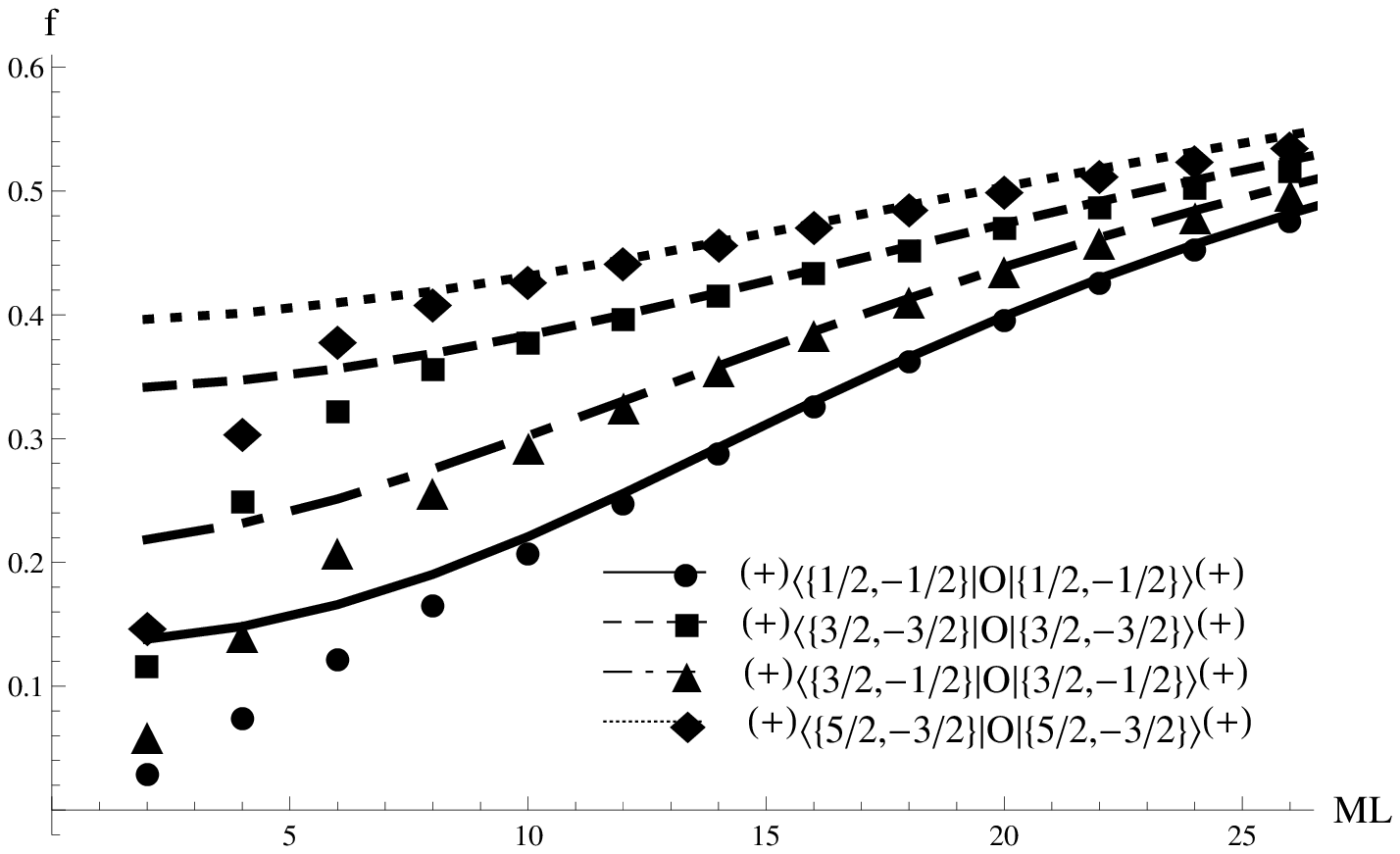} & $\quad$ & \includegraphics[scale=0.55]{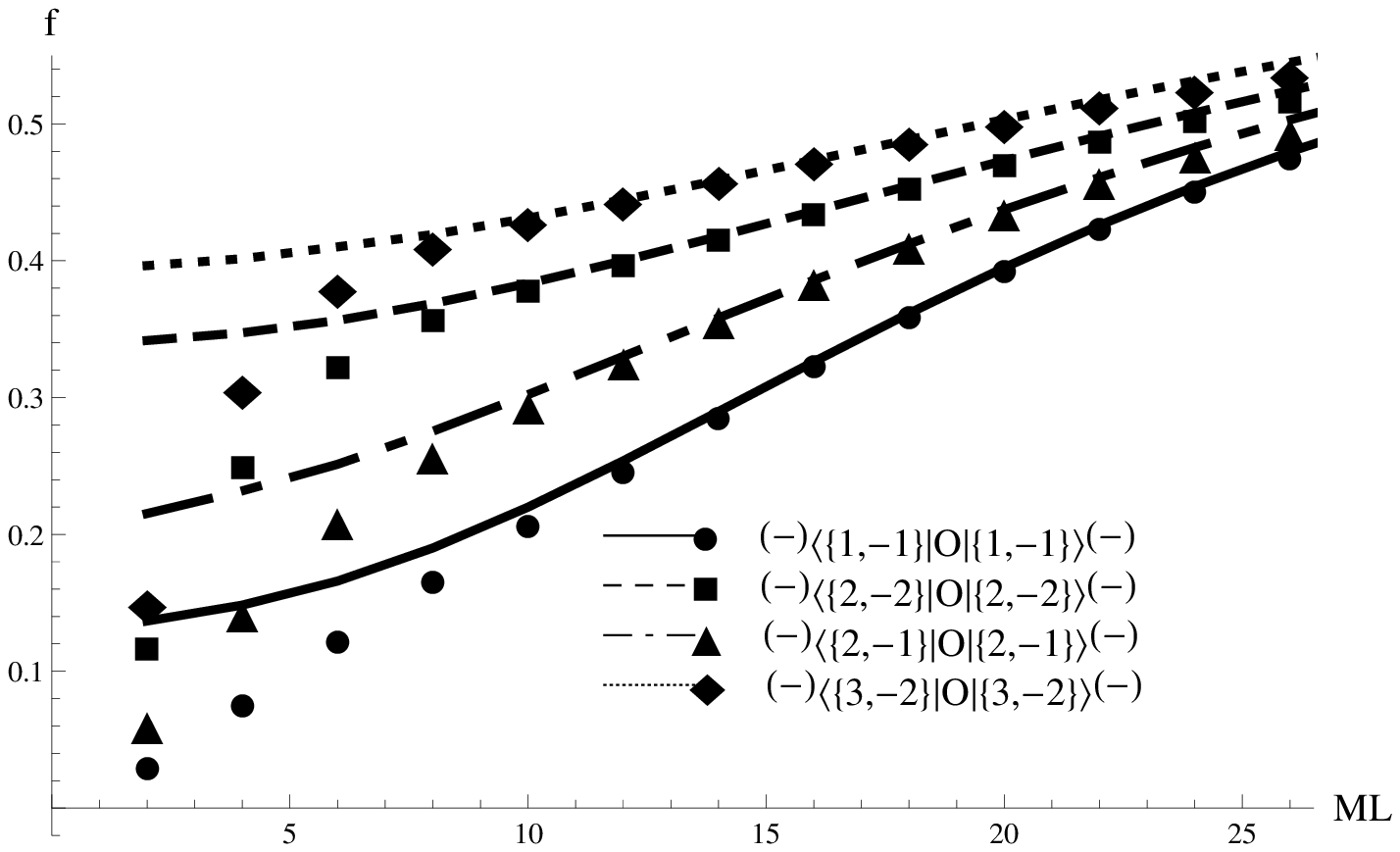}\tabularnewline
$R=1.7$, even states &  & $R=1.7$, odd states\tabularnewline
\end{tabular}
\par\end{center}

\begin{center}
\begin{tabular}{ccc}
\includegraphics[scale=0.55]{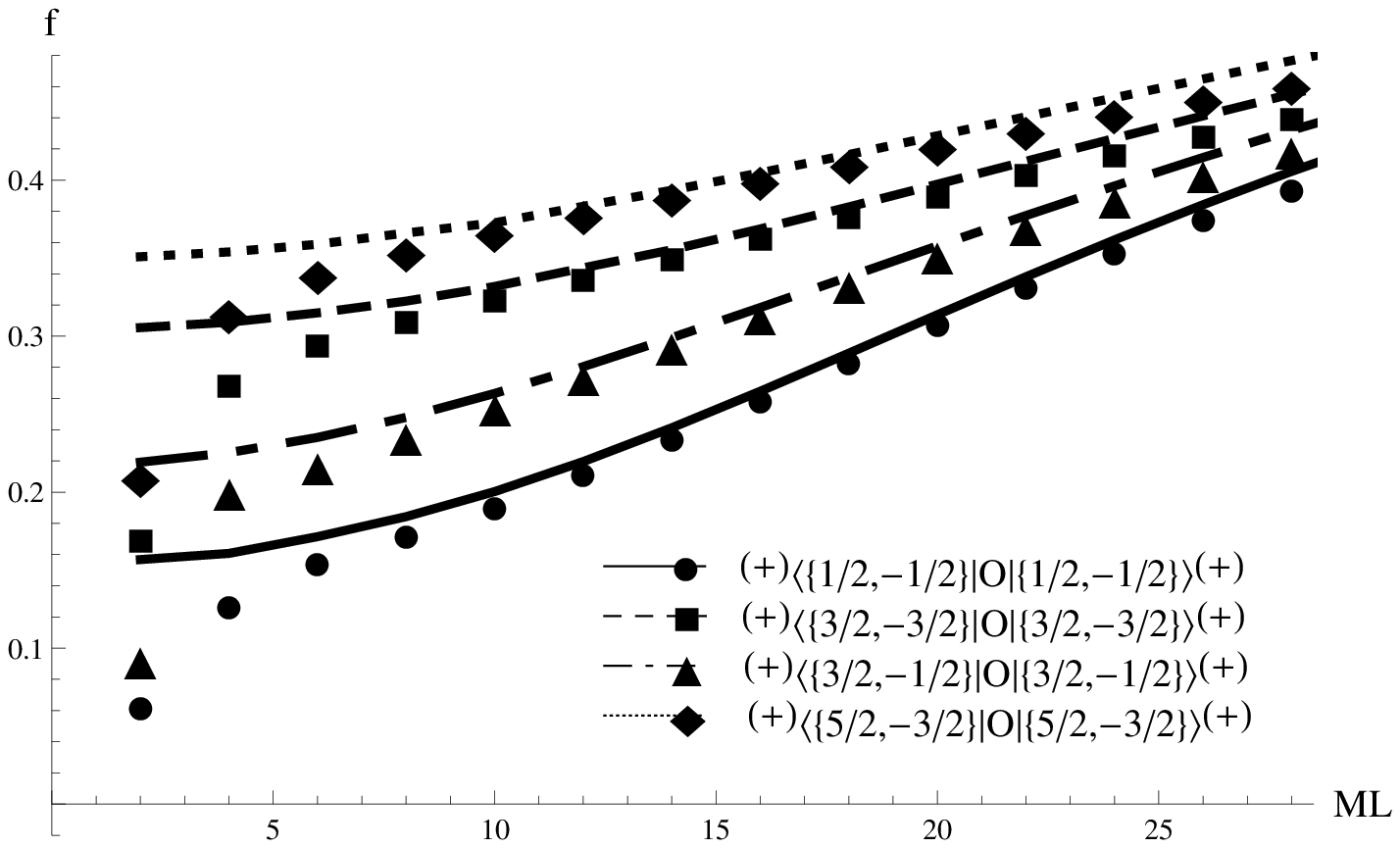} & $\quad$ & \includegraphics[scale=0.55]{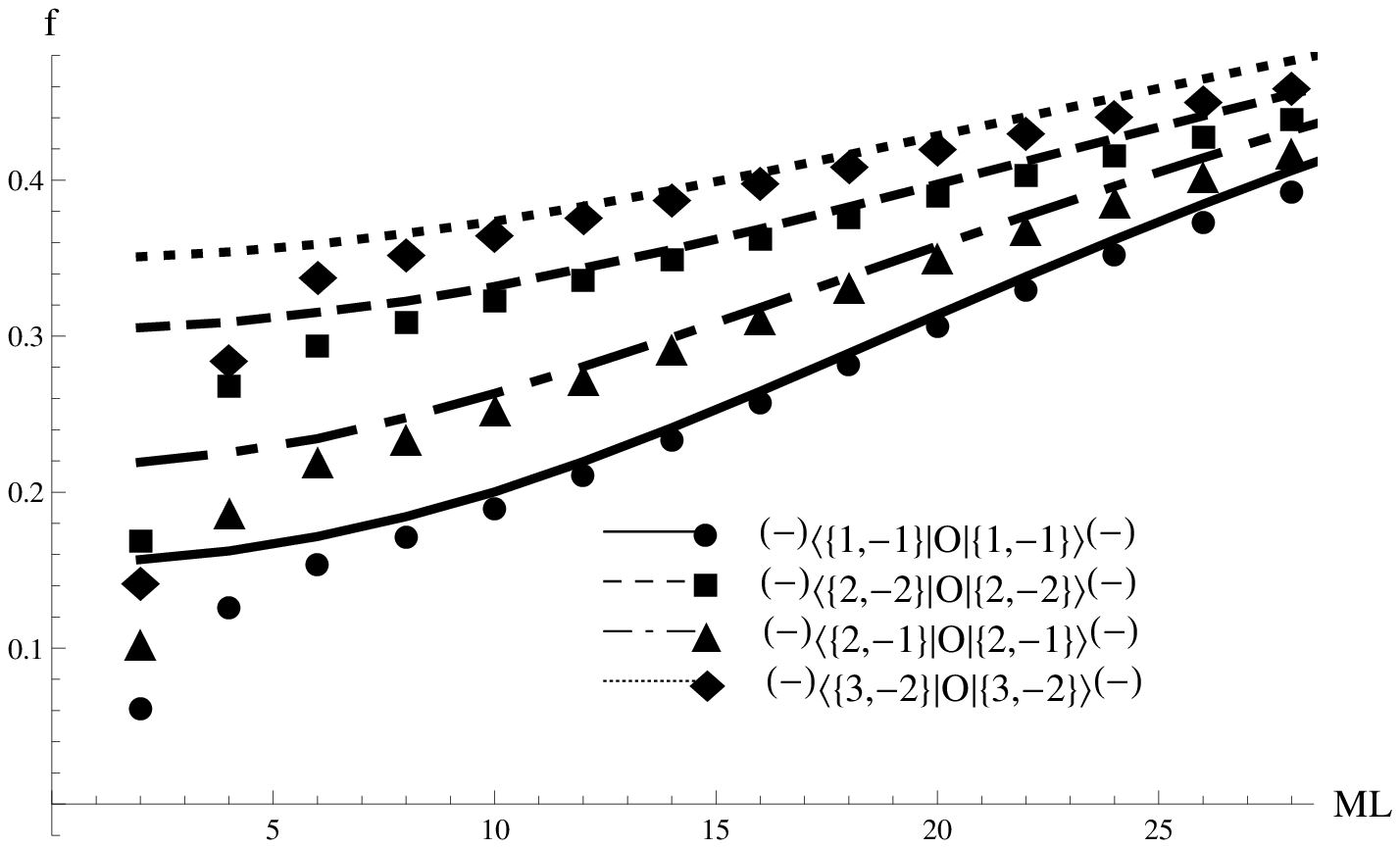}\tabularnewline
$R=1.9$, even states &  & $R=1.9$, odd states\tabularnewline
\end{tabular}
\par\end{center}

\caption{\label{fig:nondiag}Diagonal matrix elements in even and odd soliton-antisoliton states}\end{figure*}

Now we turn to the case of non-diagonal scattering. In Fig. \ref{fig:nondiag}
a large number of comparative data is shown for diagonal matrix elements
in the $\mathcal{Q}=0$ sector and compactification radii $R=1.5$,
$1.7$ and $1.9$ (corresponding to $\xi=2/7$, $50/239$ and $50/311$,
respectively). All the NRG-TCSA data are improved by the extrapolation
(\ref{eq:firstorder_rg}) using measurements up to $e_{cut}=31$ (approx.
180 thousand states). The agreement is again very convincing. Apart
from the known systematic errors the only other deviation is the breakdown
of the extrapolation in the case of higher lying states for $R=1.5$
in large volume. This is not so surprising as the truncation errors
are expected to be the largest in this case: as previously stated,
they are observed to grow when $\xi$ is increased (i.e. $R$ is decreased)
and are also expected to be larger for states higher in the spectrum. 

\section{Conclusions\label{sec:Conclusions}}

In this paper we discussed diagonal matrix elements between multi-soliton
states in sine-Gordon theory. Because of the strong cutoff dependence
of diagonal matrix elements observed in previous works \cite{Feher:2011aa,Feher:2011fn},
we first had to improve the TCSA method by taking into account more
states from the conformal Hilbert space. This was accomplished by
implementing the numerical renormalization group method proposed by
Konik and Adamov \cite{Konik:2007cb}, and then improving the resulting
NRG-TCSA data further by cutoff extrapolation, following the ideas
of the TCSA renormalization group introduced in \cite{Feverati:2006ni,Konik:2007cb}. 

The theoretical description for diagonal matrix elements in states
with diagonal scattering was known from \cite{Pozsgay:2007gx}, and
so we could use expectation values computed in two-soliton states
and three-soliton states, with all solitons having positive topological
charge, to verify the accuracy of the method. In this way we demonstrated
conclusively that the discrepancies observed previously \cite{Feher:2011fn}
were really due to cutoff effects. In fact, in our previous work we
did not even attempt to show data for diagonal matrix elements with
$\mathcal{Q}=+3$ three-soliton states because of the large inaccuracy,
but the NRG-TCSA method allowed for a satisfactory comparison of these
as well.

The main theoretical result of this paper is a conjecture for diagonal
matrix elements in states with non-diagonal scattering, formulated
in Section \ref{sec:A-conjecture-for-disconnected-parts}. For the
already known case of diagonal scattering, this conjecture reduces
to the well established formula (\ref{eq:diaggenrule}), first obtained
in \cite{Pozsgay:2007gx}. Therefore we verified the first nontrivial
case, which was diagonal matrix elements in $\mathcal{Q}=0$ two-soliton
states, i.e. states containing a soliton and an antisoliton. The conjecture
did fit the data in a very convincing way. 

It would be interesting to provide further verification with $\mathcal{Q}=+1$
three-soliton states; we tried to identify such states in the spectrum,
but due to the density of the spectrum and the exponential finite
size corrections for small volume, this proved elusive so far. There
are certain possibilities to improve this situation: using the exact
NLIE description \cite{Destri:1992qk,Destri:1994bv,Destri:1997yz,Feverati:1998dt,Feverati:1998uz}
of the finite volume levels instead of the approximation provided
by the Bethe-Yang equations (\ref{eq:betheyang_general}), and implementing
further improvements to the TCSA method. However, this is outside
the scope of this work, and we hope to return to this problem at some
point in the future.

Nevertheless, even by restricting ourselves to the results for which
we have sufficient numerical evidence, we have a complete description
of all finite volume form factors below the three-soliton threshold.
This makes it possible to evaluate the one-point and two-point functions
including all corrections below the three-soliton threshold using
the methods developed in \cite{Pozsgay:2007gx} and in \cite{Pozsgay:2010cr},
respectively. We must stress that while here we considered sine-Gordon
theory, the results are expected to be valid for general integrable
models, such as the $O(3)$ nonlinear $\sigma$ model, and are therefore
potentially applicable to a range of field theoretic and condensed
matter problems.
\begin{acknowledgments}
We are grateful to G. Watts for details on the TCSA renormalization
group, especially regarding the correct value of the exponent characterizing
the cut-off dependence. We also acknowledge very useful discussions
with Balázs Pozsgay. GT was partially supported by the Hungarian OTKA
grants K75172 and K81461.
\end{acknowledgments}
\bibliographystyle{utphys}
\bibliography{nrg}

\end{document}